\documentclass[twocolumn,showpacs,aps]{revtex4}

\usepackage{graphicx}
\usepackage{dcolumn}
\usepackage{bm}

\begin{document}

\begin{titlepage}

\title{Unusual vortex structure in ultrathin  Pb(Zr$_{0.5}$Ti$_{0.5}$)O$_3$
films }
\author{Zhongqing Wu, Ningdong Huang, Zhirong Liu, Jian Wu,
Wenhui Duan\footnote{Author to whom any correspondence should be
addressed}, and Bing-Lin Gu} \affiliation{Center for Advanced
Study and Department of Physics, Tsinghua University, Beijing
100084, People's Republic of China}
\date{\today}

\begin{abstract}
Using a first-principles-based approach, we determine the
ferroelectric pattern in PbZr$_{0.5}$Ti$_{0.5}$O$_3$ ultrathin
film. It is found that vortex stripes are formed in the system and
they are responsible for the 180$^\circ$ domains observed. When a
local external field is exerted, the vortex stripe transforms into
the vortex loop structure, which leads to the formation of a
smaller domain with the polarization antiparallel to the field in
the center of the field region. This may provide a convenient way
to manipulate nanodomains in thin films.

\end{abstract}

 \maketitle

\draft

\vspace{2mm}

\end{titlepage}

During the past decade, the extensive research on ferroelectric
thin films has been carried out not only due to their promising
applications in microelectronics such as nonvolatile random access
memories \cite{sco,was,ahn} but also due to their novel physical
properties. Ferroelectricity is a collective phenomenon resulting
from the delicate balance between the long-range dipole
interaction and short-range covalent interaction. In
nanostructures such as ultrathin films, both long-range dipole and
short-range covalent interactions and their balance are varied in
regard to those in the bulk. Therefore, it is commonly believed
that the nanostructure will have some physical properties
distinctly different from those of bulk materials, e.g., the size
effects that the ferroelectricity would disappear below a critical
size\cite{bun,str1,wan,gho,wu,jun}. Recently, an unusual atomic
off-center displacements vortex pattern in BaTiO$_3$ nanodot was
revealed in simulation\cite{fu}. In ferroelectric ultrathin films,
the periodic 180$^\circ$ stripe domains have been observed with
X-ray study \cite{str,str1} and simulated using a first-principles
based approach\cite{wu,bel,Tin}.

Generally, the ferroelectric nanodomain is read or written with a
metallic tip of some kinds of microscope such as atomic force
microscope.  Therefore, it is important to clarify how the
ferroelectric pattern of ultrathin films is influenced by the
local field produced by the tip. Herein, we conducted a
first-principles-based computer simulation to investigate this
problem. It is found that a polarization vortex loop is induced
under a local external field, and the polarizations in the center
of the field region are antiparallel with the field. This unusual
phenomenon may be helpful in manipulating the nano-structure in
the ultrathin film.

The system we investigated is the  disorder
Pb(Zr$_{0.5}$Ti$_{0.5}$)O$_3$ (PZT) thin films.
We adopt the effective Hamiltonian of PZT alloys proposed by
Bellaiche, Garcia and Vanderbilt \cite{LB1}, which is derived from
first-principles calculations, to predict the properties of the
system by Monte Carlo simulation. The method can
provide the microscopic information about the atomic off-center
displacements and therefore is especially suitable for investigating
the film without any charge screening.
In this scheme, the total
energy $E$ is written as the sum of an average energy and a local
energy as \cite{LB1,LB2}
\begin{eqnarray}
E(\{\mathbf{u}_{i}\},\{\mathbf{v}_{i}\},\eta _{H},\{\sigma _{j}\})
&=&E_{
\mathrm{ave}}(\{\mathbf{u}_{i}\},\{\mathbf{v}_{i}\},\eta _{H})  \nonumber \\
&&+E_{\mathrm{loc}}(\{\mathbf{u}_{i}\},\{\mathbf{v}_{i}\},\{\sigma
_{j}\}),
\end{eqnarray}
where $\mathbf{u}_{i}$ is the local soft mode in the $i$-th unit
cell and associated with the local electrical dipoles
$\mathbf{P}_{i}=Z^*\mathbf{u}_{i}$ (where $Z^*$ is the effective
charge of the local mode), $\mathbf{v}_{i}$ is the dimensionless
local displacement\cite{WZ}, $\eta _{H}$ is the homogeneous strain
tensors and $\sigma _{j}=\pm 1$ represents the presence of a Zr or
Ti atom at the $j$-th lattice site. All the parameters of the
Hamiltonian are derived from the first principle calculations and
are given in references \cite{LB1,LB2}. In our simulations, we do
not include the external term of surface effect proposed by Fu and
Bellaiche while simulating nanoscopic structures, since they have
demonstrated that the term has almost no effect on the
polarization pattern\cite{fu,note}. The fact that the calculated
critical thickness\cite{wu} is well consistent with the X-ray
study\cite{str1} further justifies our above treatment.

PZT film is surrounded by vacuum. To efficiently calculate the
long-range dipole-dipole interaction energy in thin films which
lack the periodicity in the out-of-plane direction, we adopt the
corrected three-dimensional Eward method, whose validity has been
verified analytically by Br\'{o}dka and Grzybowski \cite{bro}. In
that scheme, a small empty space (about three times of the film
thickness) introduced in the simulation box to surround the film
can lead to very well converged results. The $z$ axis ([001]
direction) is taken along the growth direction of the film, and
the $x$ and $y$  axes are chosen to be along the pseudocubic [100]
and [010] directions.  The
influence of the substrate is imposed by confining the homogeneous
in-plane strain. Namely, $\eta_1=\eta_2=2\%$ and $\eta_6=0$. The
temperature of the simulation is 50 K, corresponding to a rescaled
experimental temperature of 30 K\cite{LB1}.

We first investigate in detail the polarization pattern in the
ultrathin film under zero field, which is helpful for clarifying
the effect of local electric field. Our previous work demonstrated
the existence of the periodic out-of-plane 180$^\circ$ stripe
domains in the films under $2\%$ compressive strain\cite{wu}. Now
we further find that the stripes exhibit unusual microscopic
structures. An example of a $15 \times 15\times 4$ supercell is
given in Fig. 1. A special vortex structure can be clearly
observed on the $x-z$ plane [Fig. 1(a)]. The clockwise and
anticlockwise vortex appear alternately. This vortex structure
extends along the $y$ axis as vortex stripe. One of the outcomes
of the vortex stripe is that the local mode perpendicular to the
stripe, $u_x$, also forms the stripe structure in each $x-y$ plane
[see Fig. 1(b) for $z=1$ plane]. The average effect of vortex
structures directly result in 180$^\circ$ out-of-plane
polarization stripe domains. The out-of-plane polarization stripe
domain wall is located at the center of vortex. Therefore each
out-of-plane polarization domain does not superpose with the
vortex stripe but strides across two neighboring vortex stripes.
The film thickness and the stripe width just denote two
characteristic sizes of the vortex. Then we can understand
intuitively why the stripe domain period increases with the film
thickness, since the vortex expands with increasing film
thickness. While the film thickness is of 4 unit cell (1.6 nm),
the simulated stripe period is 2.8 nm, which is quantitatively
well consistent with the experimental measurement\cite{str}.

\begin{figure}[tbp]
\includegraphics[width=8.5cm]{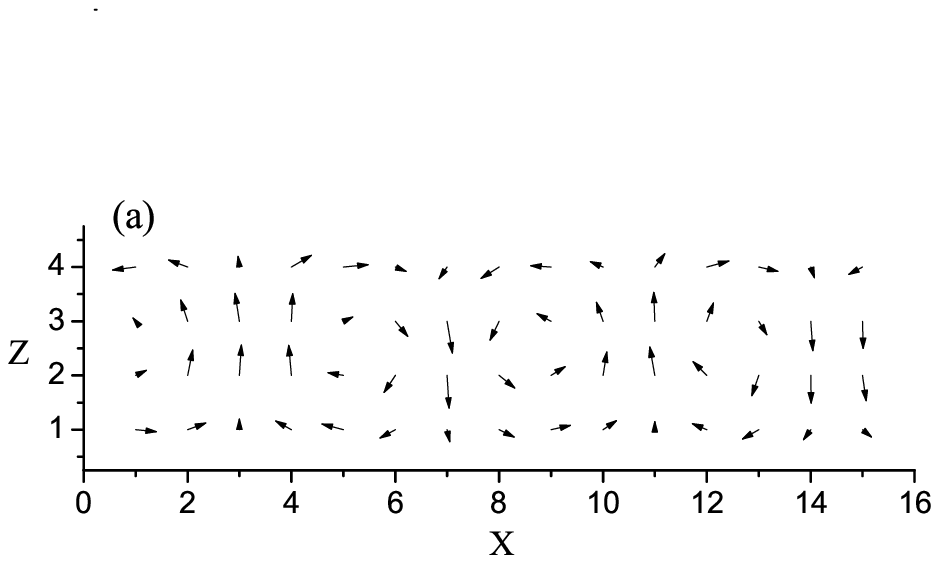}

\includegraphics[width=8.5cm]{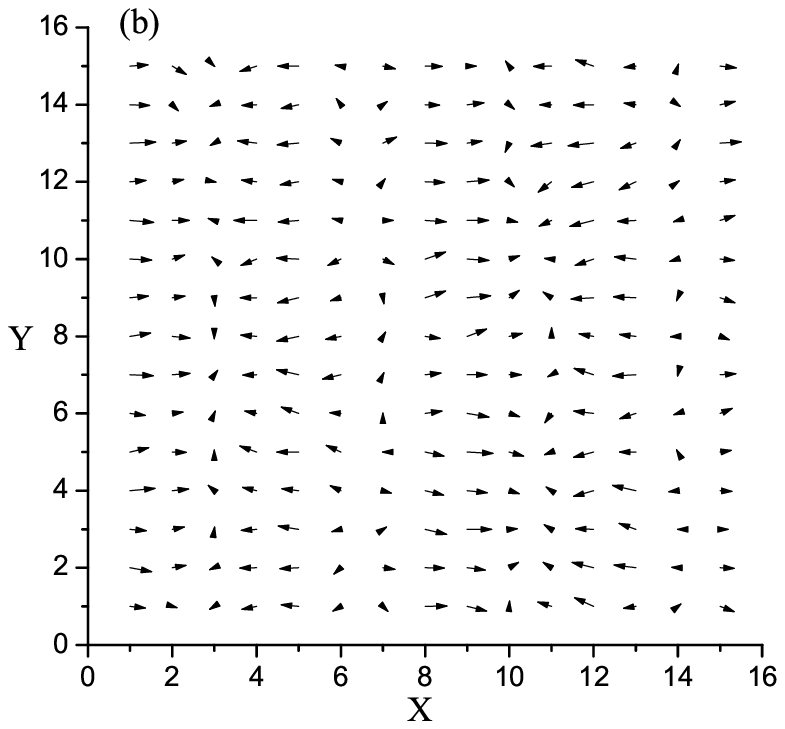}
\caption{Local-mode displacement, $\mathbf{u}_{i}$, of the cell in
(a) $y=8$ plane, (b) $z=1$ plane of $15\times15\times 4$
supercell. The arrows give the directions of these displacements,
projected on corresponding planes and the arrow length indicates
the projected magnitude. }
\end{figure}

The above vortex structure comes from the delicate balance between
the long-range dipole interaction and short-range covalent
interaction. For the dipole-dipole interaction, the alignment of
head to end is favorable in energy, while the head to head is
unfavorable. Due to the existence of the vacuum and the absence of
the screening effect, the surface dipoles tend to align parallel
with the surface to stabilize the dipole-dipole interaction. The
short-range covalent interaction supports the dipoles to change
smoothly. In the internal layers, because of the large compressive
strain, the local mode tends to align along the normal direction.
The vortex structure meet all the above demand, and, with
periodical arrangement, effectively eliminate the depolarization
field.

Now we turn our attention to the influence of the local field on
the microscopic structure. In our simulation, the local field
produced by the tip is assumed to exert in a smaller region,
$N_E\times N_E\times 4$, of the supercell. Due to the periodic
boundary condition adopted in the film plane, the local field will
appear periodically, which is similar to the field created by the
array of the tips. Generally, we can expect that a ferroelectric
nanodomain will be formed when a local field is exerted. However,
to our surprise, the actual situation is not so simple.  Fig. 2
depicts the out-of-plane local mode (namely polarization) of the
film with $E=2\times10^7$ V/m and $N_E=10$. Although the field
aligns most out-of-plane polarizations in the field region along
the field direction, the polarizations near the center of the
field region are opposite with the field. The smaller region where
out-of-plane polarization is antiparallel with the field is
approximately round in shape.

\begin{figure}[tbp]
\includegraphics[width=8.5cm]{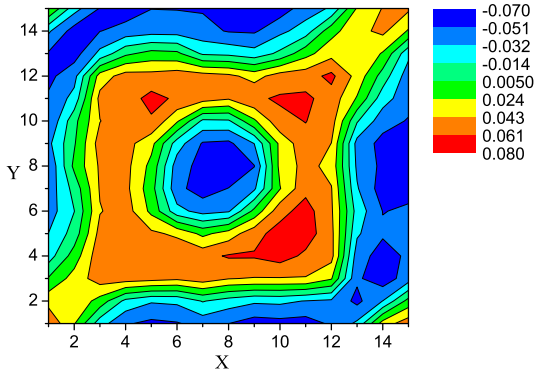}
\caption{(color online) The out-of-plane local modes of
$15\times15\times 4$ supercell. A local field $E=2\times10^7$ V/m
is exerted on a smaller region of $10\times 10\times 4$ (namely
$N_E=10$).}
\end{figure}

More details of the polarization pattern are given in Fig.~3.
Although the clockwise and anticlockwise vortexes still appear
alternately in the $x-z$ plane [Fig.~3(a)], the straight stripe in
the $x-y$ plane is destroyed by the local field and a vortex loop
appears in the field region as shown in Fig.~3(b). Our detailed
analysis shows that the vortex loop structure could be regarded as
being formed by the rotation of the inner vortex around the
$oo^\prime$ axis [Fig.~3(a)]. Fig.~3(b) supports this conclusion
by means of the facts that the boundary line between the inner
vortex and the outer vortex is roughly round (the dashed circle in
the graphics) and the local modes inside the circle are almost
radialized. Similar proof is also observed in Fig.~2. Therefore,
our simulations demonstrate that the local field can result in the
transformation of the vortex straight stripe into the vortex loop.

\begin{figure}[tbp]
\includegraphics[width=8.5cm]{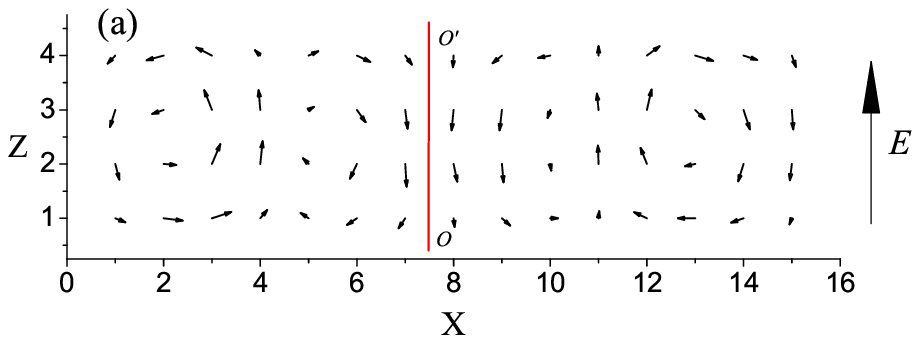}

\includegraphics[width=8.5cm]{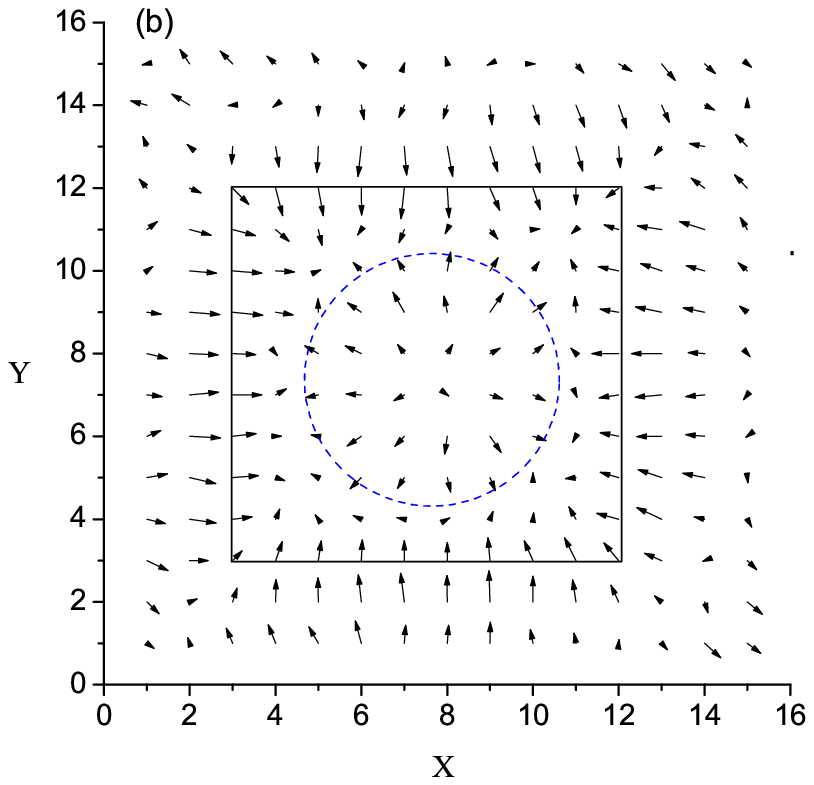}
\caption{Local-mode displacements, $\mathbf{u}_{i}$, of the cell
in (a) $y=8$ plane, (b) $z=1$ plane of $15\times15\times 4$
supercell.  The arrows give the directions of these displacements,
projected on corresponding planes and the arrow length indicates
the projected magnitude. A local field $E=2\times10^7$ V/m along
the $z$ axis is exerted on a smaller region $10\times 10\times 4$
denoted by solid line. The dash line shows the boundary of the
inner vortex and the outer vortex. }
\end{figure}

The above unusual phenomena are closely related with the character
of the stripe. On the one hand, the shape of vortex is rather
stable. The stripe width is determined only by the film thickness
and it is almost independent of the temperature and compressive
strain until the stripe disappears. The influence of the global
external field (i.e., the field is exerted on the whole supercell)
on the stripe width is also considerably small. Our simulations
indicate that the stripe width does not alter appreciably after
applying a very large external field ($10^8$ V/m) to the
supercell. All of these results clearly demonstrate that the shape
of vortex is very stable. On the other hand, the stripe structure
composed of the vortex is considerably flexible. Our simulations
show that besides the straight stripe mentioned above, many kinds
of bent stripes also correspond to local minimum of the energy.
Although the internal energy of these bent stripes are larger than
that of the straight stripe, the energy difference is very small.
Therefore we can conveniently trap some bent stripe by applying
and removing the local field with different configuration. For
example, similar vortex loop structure with Fig. 2 can be trapped
when applying and then removing the local field. The corresponding
internal energy is larger than that of the straight stripe by only
about 1 meV /5 atoms. The diameter of the inner domain in the
vortex loop structure is smaller than the stripe period $\tau$
($\tau\approx 7$ in the current case). Therefore, when $N_E$
(denoting the field region of the external local field) exceeds
$\tau$, no configuration of polarization can make the whole field
region occupied by the parallel phase. Alternately, a loop
structure with antiparallel polarization in the center, as that in
Figs.~2 and 3, is the optimum scheme to have as much parallel
polarization as possible in the action range of the local field.
It is the reason of the appearance of the central antiparallel
domain. When the external field range $N_E$ further increases, the
number of the loop stripe in the structure will increase due to
the fact that the width of the loop stripe is almost fixed, and a
parallel domain (data not shown) may appear in the center.

Our simulations indicate that the same phenomena also appear in
thicker films (such as the films with the thickness of 6 and 8
unit-cells). Furthermore, the radius of central anti-parallel
domain is only determined by the film thickness. The influence of
the size of the field region and in-plane size of the supercell on
the radius of anti-parallel domain could be overlooked. The radius
of anti-parallel domain is a little larger than half of the stripe
width, and is much smaller than the size of the field region
(denoted by $N_E$). Therefore, our results imply that we can write
a naondomain whose size is far smaller than the radius of tip.
This unique character may have potential application in improving
the storage densities of ferroelectric memory. In general, in
order to improve the storage density, we need to decrease the
radius of the tip since the size of tip is a very important factor
to determine the lower limits of the radius of nanodomain.
However, it is obvious that decreasing of the tip radius has its
own limit below which it would be very difficult to further reduce
the tip. Our simulation suggests a possible approach to break
through the limit imposed by the radius of tip. According to our
calculated results and the fact that the stripe width of the
100-nm-thick PbTiO$_3$ film is about 10 nm by X-ray study
\cite{str}, we can estimate that the radius of anti-parallel
domain in an 100-nm-thick PbTiO$_3$ film is about 5 nm,
corresponding to a considerably small dot.

It should be noted that the formation of the vortex structure such
as the vortex stripe and loop is closely related with the electric
boundary condition adopted in our simulations (i.e., no surface
charge induced by polarization is screened). With enhancing
screening of surface charges, the vortex stripe domains will
transform into monodomain\cite{bel}. In fact, the perfect
screening is not achieved even under short circuit boundary
conditions in ultrathin film. For example, a sizeable
depolarization field is quantified by Junquera and
Ghosez\cite{jun} through the first principles calculation. Such a
field was recently suggested to explain experimental changes of
the coercive fields with the film thickness\cite{daw}. The effect
of the screening became weaker when the top electrode has been
replaced by the metallic tip, which is just the case of writing
the nanodomain in ferroelectric films. A larger depolarization
field will be formed in those ferroelectric films. Therefore, the
phenomena that the direction of the polarization is antiparallel
with the field near the center of the field region will be
promisingly observed in perovskite ferroelectric film.

Similar anti-parallel poling reversal phenomenon has been reported
by Abplanalp {\it et al.}\cite{abp} and Morita {\it et
al.}\cite{mor}. However, this phenomenon is not induced by the
depolarization field, and has different microscopic origin from
ours. In Abplanalp {\it et al.}'s study\cite{abp}, the
anti-parallel reversal is ferroelastoelectric switching and is
achieved in BaTiO$_3$ thin film by simultaneously applying
electric field and compressive stress with the tip of a scanning
force microscope. In Morita {\it et al.}'s study\cite{mor}, the
depolarization field should be small since the film can be poled
into monodomain. Although the reason of anti-parallel poling
reversal is still unclear, it is suggested that this phenomenon is
related to the mechanical force imposed by the tip. Furthermore,
the anti-parallel poling appears only after the electric field is
removed, which is different from our results that the
anti-parallel polarization starts to appear under the local field.

In summary, the ferroelectricity of ultrathin PZT films without
any charge screening has been investigated with the Monte Carlo
simulations based on a first-principles-derived Hamiltonian. The
off-center displacement in the film exhibits vortex stripes with
unusual characters. On the one hand, the shape of vortex is rather
stable. On the other hand, the stripe structure composed of the
vortex is considerably flexible and can be easily manipulated. The
unusual characters of the vortex stripe lead to an interesting
phenomenon. The local field can transform the vortex straight
stripe into the vortex loop structure, which results in the
formation of a round-shape domain with the polarization
antiparallel with the field near the center of the field region.
The area of the anti-parallel domain is much smaller than the area
of the field region. This implies a way to write a very small
nanodomain with the general tip, which might have some potential
application in improving the storage densities of ferroelectric
memory.

This work was supported by State Key Program of Basic Research
Development of China (Grant No. TG2000067108), the National
Natural Science Foundation of China (Grant No. 10325415), Ministry
of Education of China, and China Postdoctoral Science Foundation.


\begin{thebibliography}{}
\bibitem{note} Although the above effective Hamiltonian was developed for
bulk PZT, the surface effect is partially and implicitly
considered during the simulations on the films by i) including the
effect of the broken bonds in the surfaces, which is substaintial
for the surface energy; and by ii) relaxing the local soft mode
vectors and inhomogeneous strain, which partially refelects the
effect of surface relaxation.
\end{thebibliography}
\end{document}